\documentclass[twocolumn,prb,showpacs,preprintnumbers,amsmath,amssymb,superscriptaddress]{revtex4}

\usepackage{upgreek}
\usepackage{graphicx}
\usepackage{dcolumn}
\usepackage{bm}

\begin{document}

\preprint{arXiv preprint}

\title{Coupling an epitaxial quantum dot to a fiber-based external-mirror microcavity}

\author{Andreas Muller}
 \email{andreas.muller@nist.gov}
\affiliation{%
Joint Quantum Institute, National Institute of Standards and Technology and University of Maryland, Gaithersburg, MD}
\author{Edward B. Flagg}%
\affiliation{%
Joint Quantum Institute, National Institute of Standards and Technology and University of Maryland, Gaithersburg, MD}
\author{Michael Metcalfe}%
\affiliation{%
Joint Quantum Institute, National Institute of Standards and Technology and University of Maryland, Gaithersburg, MD}
\author{John Lawall}
\affiliation{Atomic Physics Division, National Institute of Standards and Technology, Gaithersburg, MD
}%
\author{Glenn S. Solomon}%
 \email{glenn.solomon@nist.gov}
\affiliation{%
Joint Quantum Institute, National Institute of Standards and Technology and University of Maryland, Gaithersburg, MD}
\affiliation{Atomic Physics Division, National Institute of Standards and Technology, Gaithersburg, MD
}%

\date{\today}

\begin{abstract}
We report the coupling of individual InAs quantum dots (QDs) to an external-mirror microcavity. The external mirror is bonded to a fiber and positioned above a semiconductor sample consisting of a QD-containing GaAs layer on top of a distributed Bragg reflector (DBR). This open cavity can be rapidly tuned with a piezoelectric actuator without negatively affecting the QD linewidth. A mirror radius of curvature of 42 $\upmu$m and a cavity length of 10 $\upmu$m enable good mode-matching and thus high collection efficiency directly into the fiber. With an improved finesse this system may enter the strong coupling regime.\end{abstract}

\maketitle

Within the past decade, single-emitter based cavity quantum electrodynamics (QED), long restricted to atoms and ions, has become a major topic of solid-state physics research, taking advantage of ``artificial atoms" such as semiconductor quantum dots (QDs),\cite{solomon2001sms,reithmaier2004scs,yoshie2004vrs, peter2005eps,srinivasan2008icd} Cooper-pair boxes,\cite{fink2008cjc} and impurity centers.\cite{park2006cqd} In the near-infrared region, the strong coupling between a single QD and a single cavity mode has been demonstrated using micropillar, \cite{reithmaier2004scs} photonic-crystal defect,\cite{yoshie2004vrs} and microdisk\cite{peter2005eps,srinivasan2008icd} cavities, and several other promising designs are being pursued.\cite{rakher2009emm,muller2006saa} These approaches all have one thing in common: the cavity is fabricated permanently around the QD, making very small mode volume, $V$, and therefore very high cavity-emitter coupling rates, $g\propto1/\sqrt{V}$, possible. However, these benefits come at the cost of quality factor, $Q$, and severe restrictions on cavity tunability. For example, sample heating is the most common method used to bring a QD transition in resonance with a single cavity mode. But besides being a very slow method, the tuning range is limited at elevated temperatures by phonon scattering that increases the QD linewidth, $\gamma$. The proximity of QDs to etched surfaces is also known to increase $\gamma$.\cite{gerard1998ese} While these drawbacks have not hindered initial proof-of-concept demonstrations of weak and strong coupling, they make further progress very difficult, and require heroic efforts for spatial and spectral tuning.\cite{dousse2008clm} These difficulties likely contribute to our limited understanding of off-resonant QD/cavity coupling, \cite{hennessy2007qns} and prevent the realization of advanced cavity-QED phenomena such as single QD lasers and multiphoton interactions. \cite{fink2008cjc}

Because of the record-high finesse of traditional macroscopic Fabry-P\'erot cavities and the rapid development of micromirror technologies, an external-mirror open cavity design offers an attractive way forward. Micromirrors have been used for many years in atomic physics, either fabricated on, or bonded to a fiber, \cite{steinmetz2006asf, trupke2005mhf} achieving mode-volumes as low as 600 $\upmu m^3$ and a finesse as high as 37 000.\cite{colombe2007saf} For InAs QDs with dipole moments of $d\approx$ 10 Debye coupled to the single mode of such a cavity, this translates into a coupling constant $g/2\pi$ of several GHz, exceeding the QD linewidth, $\gamma/2\pi\approx$ 1 GHz, as well as the cavity linewidth, thus reaching the strong coupling regime.

\begin{figure}[b]
\includegraphics[width=3.4in]{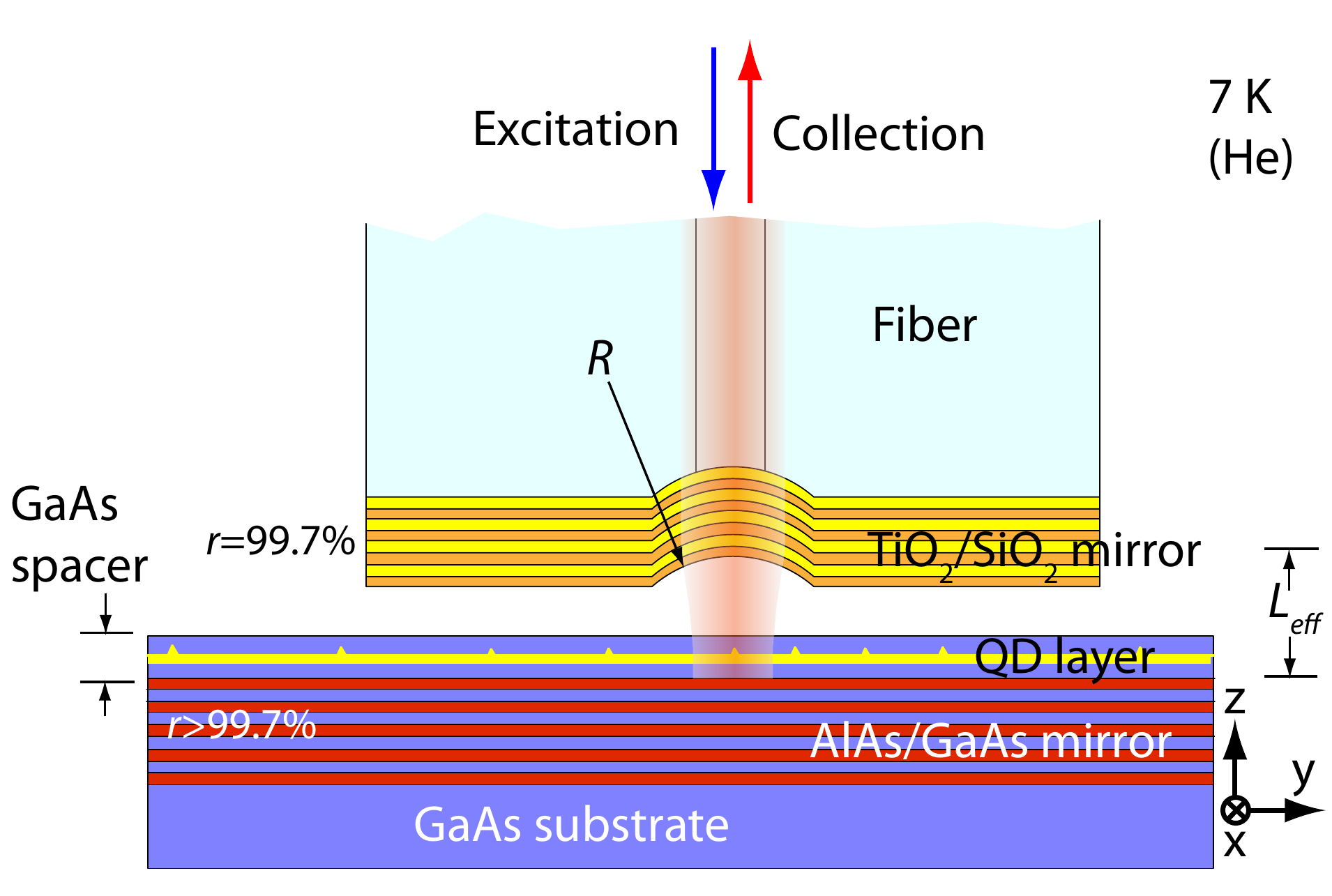}
\caption{\label{fig1} (Color online) Schematic of external-mirror microcavity. A non-resonant pump laser is introduced into the fiber to excite the QDs, and the QD emission is collected into the same mode. The distance between the sample and the curved mirror (radius $R$) is precisely controlled with a piezo actuator to tune the cavity.}
\end{figure}

Motivated by these possibilities we have implemented an open cavity that consists of a fiber-bonded micromirror of moderate reflectivity, and a planar distributed Bragg reflector (DBR) that contains the QDs. We show that it results in a stable cavity that is nearly mode-matched to the single mode fiber and is precisely tunable by a voltage applied to a piezo actuator. Employing the fiber eliminates alignment difficulties which have perhaps made prior QD open-cavity attempts unsuccessful. \cite{cui2006ahh} Through $x-y-z$ nanopositioning of the semiconductor sample, we can maximize the spatial QD/cavity mode overlap, and macroscopically modify the cavity length, crucial for applications requiring doubly-resonant coupling, such as entangled photon sources based on exciton-biexciton emission each coupled to a different cavity mode. For a cavity length of $\approx$ 10 $\upmu$m, we collect into the fiber about 10 \% of all photons emitted by a single QD at 7 K. This collection efficiency exceeds by several orders of magnitude the collection achieved from QDs in a planar optical cavity using a high numerical aperture  (NA) microscope objective. \cite{muller2007rfs} Our open cavity approach thus also constitutes a spectroscopic tool in itself, and a promising means of entangling the state of a single photon with a QD exciton which requires the capability to collect all photons spontaneously emitted, or conversely the complete extinction of an input laser by a single exciton.\cite{vamivakas2007sef}

The semiconductor sample was grown by molecular-beam epitaxy and consists of thirty-two quarter-wave pairs of alternating layers of GaAs and AlAs, and a $4\lambda/n$-thick GaAs spacer which contains the QD layer at its center. $\lambda$ = 920 nm is the target wavelength, and $n$ is the 4 K index of refraction. The QDs are a dilute ensemble ($\approx$ 10 $\upmu m^{-2}$) of strain-induced InAs islands, emitting in the range 880 nm to 980 nm. The reflectivity of this sample was chosen to be larger than that of the external mirror, resulting in an unbalanced cavity favoring emission into the collection direction. The fabrication of the external mirror is based on a modification of a previous design \cite{steinmetz2006asf} that uses transfer coatings.\cite{OIB}  To obtain very small radii of curvature, convex microlens arrays etched into fused silica were used as a source template. The coating, consisting of a release layer and a SiO$_2$/TiO$_2$ dielectric stack, was deposited onto the template. The coating is highly reflective in the range 850 nm to 950 nm ($r$ $\approx$ 99.7 \% at 920 nm), and transmissive for $\lambda\lesssim$ 830 nm. A single mode fiber is aligned with a microlens and attached to the coating with transparent epoxy. Upon pull-back of the fiber, the coating is separated at the release layer and transfered onto the fiber as a concave mirror. The fiber is then positioned above the semiconductor sample and the entire assembly immersed into a He bath cryostat filled with He exchange gas, as depicted in Fig. 1.

To characterize the cavity we introduce a narrow linewidth continuous-wave laser into the fiber and measure dips in the reflected signal as the spacing between the concave mirror at the fiber tip and the semiconductor sample is varied with a piezoelectric actuator. Alternatively, the cavity mode spectrum can be observed in photoluminescence (PL) under strong above-band pump. For excitation we use either a HeNe laser or a picosecond pulsed Ti:sapphire laser at $\approx$ 790 nm that generate carriers non-resonantly in the GaAs matrix that surrounds the QDs. Emission from QDs at or very near a cavity resonance is collected directly into the single mode fiber. Spectral measurements of this emission are performed with a spectrometer and a liquid Nitrogen-cooled charge coupled device (CCD) detector.

Figure 2(a) shows PL mode spectra for different cavity lengths, $L_{eff}$. $L_{eff}$, calculated from the measured free-spectral-range, $FSR=c/2L_{eff}$, includes the four-wavelength QD containing spacer (optical thickness $\approx$ 3.7 $\upmu m$), the He gas gap, and the height $h\approx$ 2 $ \upmu m$ of the concave mirror. The radius of curvature of the latter is nominally $R=$ 42 $\upmu$m. In Fig. 2(a), the evolution of the spectra is shown when finely tuning the cavity length around 10.5 $\upmu$m (bottom), 20  $\upmu$m (middle), and 47 $\upmu$m (top). The $FSR$ decreases with increasing cavity length, and at $L_{eff}=$ 47 $\upmu$m the cavity is at the edge of stability, as expected when concentric ($L=R$); when $L\gtrsim47$ $\upmu$m no PL is observed.

\begin{figure}[t!]
\includegraphics[width=3.5in]{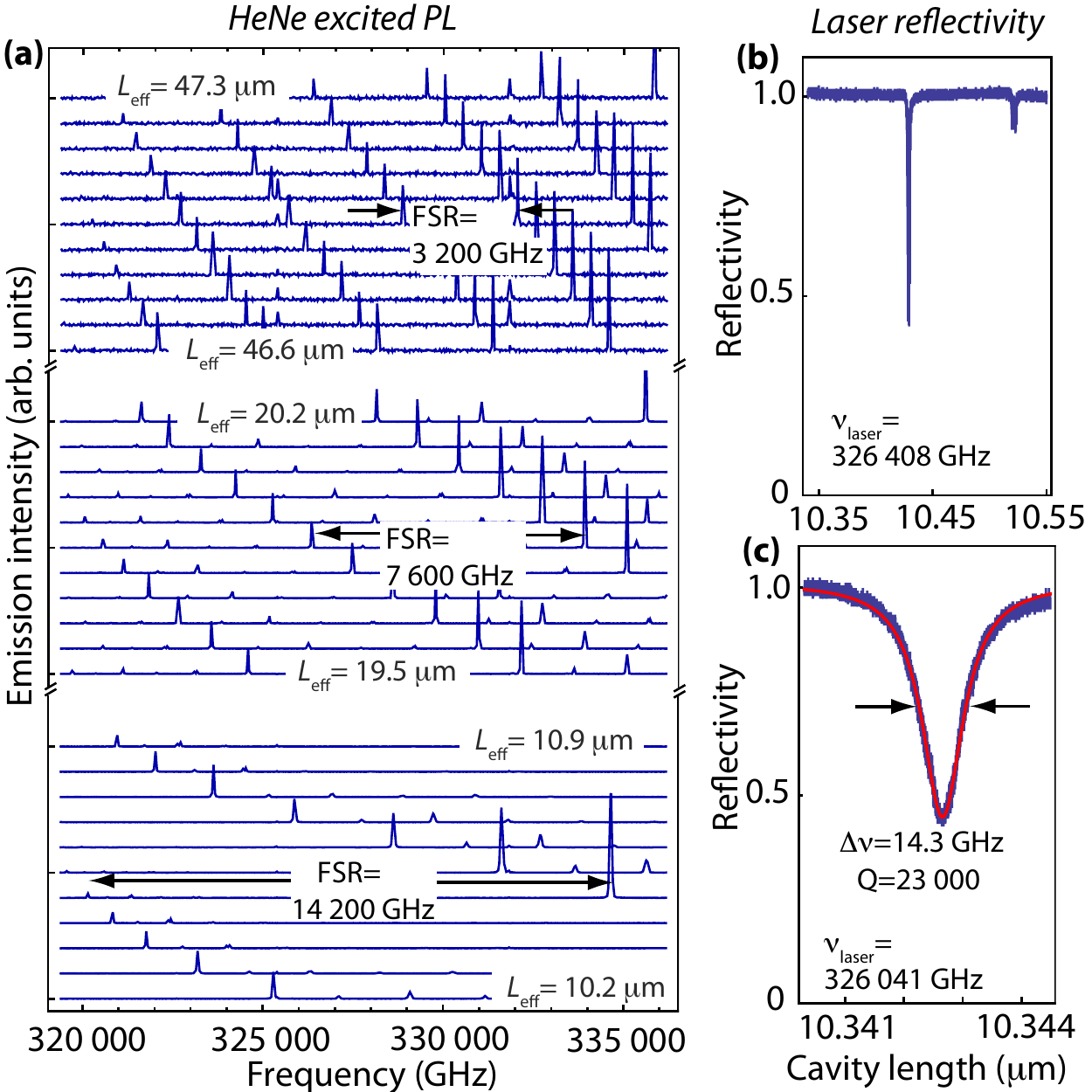}
\caption{\label{fig2} (Color online) (a) Photoluminescence (PL) spectra of the cavity with varying cavity size subject to strong HeNe laser pumping. (b, c) Reflection signal of narrow band laser as the cavity size is changed around $L\approx$ 10 $\upmu$m. A fundamental lateral cavity mode as well as the corresponding first higher-order lateral mode are seen, as in PL. The cavity $Q$ is determined by scanning the laser frequency over the resonance.}
\end{figure}

High resolution reflectivity measurements are shown in Fig. 2(b) and (c). $L_{eff}$ is scanned around $\approx$ 10.5 $\upmu$m while the reflected laser intensity is collected at fixed laser frequency. By means of changing the laser frequency, the full width at half maximum of the fundamental mode was determined to be $\Delta\nu$ = 14.3 GHz, and thus $Q$ = $\nu/\Delta\nu$ = 23 000. The cavity finesse, $\mathcal{F}$ = $FSR/\Delta\nu$, is about 1000. For this cavity size, the spot size at the waist evaluates to $w_0=\sqrt{\lambda((R-L_{eff})L_{eff})^{1/2}/\pi}$ = 2.3 $\upmu$m, and the mode volume is $V=\pi w_0^2 L_{eff}/4$= 44 $\upmu m^3$. This value is larger than those obtained for fully-integrated microcavities, but for cavity QED effects to manifest, a very high $Q$ can be advantageous over very small $V$, so long as the cavity linewidth exceeds that of the QD.\cite{steinmetz2006asf} $R$ could in principle be decreased further, and so far we have employed micromirrors with a radius as small as $R$ = 34 $\upmu$m. As in previous studies using transfer coatings, we conclude that surface roughness of the mirror currently limits the cavity Finesse. Alternate approaches for obtaining higher finesse have already been demonstrated. \cite{colombe2007saf}

Using the above cavity configuration, single QD resonances are easily measured, provided the pump laser intensity is not much larger than the QD saturation intensity. These are represented in a large area map in Fig. 3(a). Detailed spectra are shown in Fig. 3(b) and (c) for the regions circled in Fig. 3(a), labelled QD1, QD2, and QD3. When the mode is resonant with a single QD transition the QD emission is drastically increased. It is also much larger than the emission obtained from QDs in planar microcavities with a microscope objective. High-resolution measurements with a scanning Fabry-P\'erot interferometer    reveal QD spectral resonances as narrow as about 0.5 GHz [inset of Fig. 3(a)], verifying that the external mirror does not deteriorate the linewidth.

\begin{figure}[t!]
\includegraphics[width=3.5in]{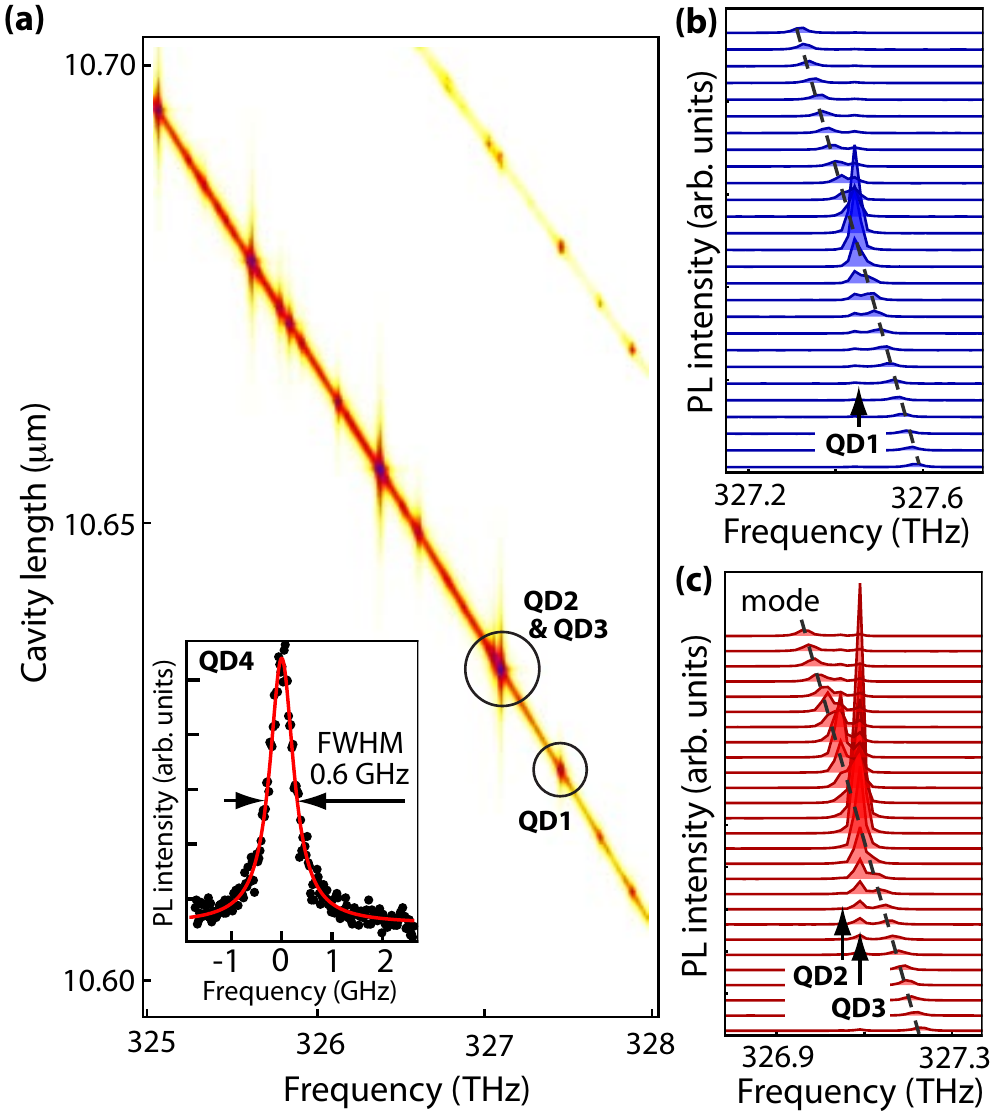}
\caption{\label{fig3} (Color online) (a) PL spectral map (high resolution) of the cavity with varying cavity length subject to moderate HeNe laser pumping. The inset shows the lineshape of the narrowest QD resonance encountered, resolved with a scanning Fabry-P\'erot interferometer. (b, c) Zoom into three QD/cavity crossings, labelled QD1, QD2, and QD3. The PL spectra, offset for clarity, were recorded for detunings ranging from about -150 GHz to +150 GHz.}
\end{figure}

We have measured a small decrease in PL lifetime on resonance compared to off-resonance, indicating the onset of cavity QED phenomena, i.e. the Purcell effect.\cite{gerard1998ese,solomon2001sms} However, the strong signal is primarily due to a good matching to the mode of the fiber, evident from the significant reflectivity dip in Fig. 2(b, c). The spot size at the curved mirror, $w_1$ = 2.7 $\upmu$m, is in fact very close to the waist accepted by the fiber (NA $\approx$ 0.11). We obtain about $10^5$ detector counts per second under pulsed non-resonant pumping at 76 MHz while $g^{(2)}(0)<0.5$ holds ($g^{(2)}(\tau)$ denotes the second-order correlation function of the emitted light). Assuming that this pumping rate is distributed equally among both bright and both dark states, and accounting for the collection efficiency of our setup (measured to be $\approx$ 14 \%) and the detector efficiency (nominally 34 \%), we calculate that as many as 10 \% of all photons emitted are collected {\em into the single mode fiber}. This figure compares well with that achieved in micropillar cavities \cite{santori2004scp} considering that no effort has been made to optimize the angle between the sample and the micromirror, and that there is room for improvement. For instance, the beam is diverging at the fiber input, so perfect mode matching will require an additional lens between the fiber and the concave mirror. Resonant measurements under excitation with a Rabi $\pi$ pulse will also be needed for a better quantification of extraction efficiency. The open cavity design is in fact well-suited for such resonant excitation because it supports waveguide modes for background-free detection of resonance fluorescence. \cite{muller2007rfs} It offers exciting opportunities for advanced quantum optical experiments that derive from strongly laser-driven effects as well as from cavity QED effects.

We acknowledge NSF support through the PFC@JQI.

\end{document}